\documentclass[manuscript]{aastex}
\usepackage{txfonts}
\usepackage{latexsym,bm}
\usepackage{lineno}
\usepackage{multirow}
\shorttitle{Gradual SEPs Calculations and Observations}
\shortauthors{Qin and Wang}

\begin{document}

\title{SIMULATIONS OF A GRADUAL SOLAR ENERGETIC PARTICLE EVENT OBSERVED BY HELIOS 
1, HELIOS 2, AND IMP 8}

\author{ Gang Qin\altaffilmark{1} and Yang Wang\altaffilmark{1}}  
\email{gqin@spaceweather.ac.cn; ywang@spaceweather.ac.cn}
\altaffiltext{1}{State Key Laboratory of Space Weather,  National Space Science Center, Chinese Academy of Sciences, Beijing 100190, China}

\begin{abstract}
In this work, a gradual solar  energetic particle (SEP) event observed 
by multi-spacecraft has been {\bf{simulated.}} 
The time profiles of  SEP fluxes accelerated by an interplanetary 
shock in the three-dimensional interplanetary space are obtained  
by  {\bf{solving numerically}} the Fokker-Planck  focused transport equation.
{\bf{The interplanetary shock is modeled as a moving source of energetic 
particles.}}
{\bf{By fitting the 1979/03/01 SEP fluxes observed by $Helios$ 1, $Helios$ 2, 
and $IMP$ 8 with our simulations}}, we obtain the best 
parameters for the shock acceleration efficiency model. 
And we also  find  that  the particle perpendicular diffusion 
coefficient with the level of $\sim 1\%-3\%$ of
parallel diffusion coefficient at $1$ AU should be included. 
{\bf{The reservoir phenomenon is reproduced in the simulations,
and the longitudinal gradient of SEP fluxes in the decay phase, which is
observed by three spacecraft at different locations,}}
is more sensitive to the shock acceleration 
efficiency parameters than that is to 
the perpendicular diffusion coefficient.

\end{abstract}


\section{INTRODUCTION}
There are two categories of solar energetic particle (SEP) events: impulsive events 
and gradual events. The impulsive events are small, last for hours, are rich in 
electrons, ${}^3$He and heavy ions, have relatively high charge states, and are 
produced by solar flares. In contrast, gradual events are large, last for days, are 
electron poor, have relatively low charge states, and are related to the shocks 
driven by interplanetary coronal mass ejections (ICMEs). 
Some large ICME driven shocks can cover a large range of solar longitudes and 
latitudes in the interplanetary space, and the observers located at different 
locations can be connected to different parts of the shocks by interplanetary 
magnetic field (IMF). Therefore, multi-spacecraft observations by, 
e.g., $Helios$ 1 and 2, usually show a {\bf{large}} difference of SEP time profiles at 
different longitudes \citep{Reames1996ApJ...466..473R, Reames1997ApJ...491..414R}. 
{\bf{In addition, in some SEP events, both of solar flare and shock can exist in 
the same event. Therefore, there are also shock-flare-mixed SEP events which can 
be identified by abundance ratio and charge state of heavier ions.}}

Multi-spacecraft observation data provide essential information to understand the 
processes of particle acceleration and transport in the heliosphere. 
Multi-spacecraft observations in the {\bf{ecliptic plane}}, e.g., by $Helios$ 1 and 
2, or at different latitudes and radial distances, e.g., by $ACE$ and $Ulysses$,
 usually show two interesting phenomena in some gradual events. 
For the first one, SEP events could be  observed by multi-spacecraft 
with a very wide spatial distribution that could be wider than the size of the 
source  \citep[e.g.,][]{Wiedenbeck2013ApJ...762...54W}. 
{\bf{This phenomenon can be explained in part as a result}} of the effect of  particle 
perpendicular diffusion, and has been investigated in detail with simulations 
\citep{Zhang2009ApJ...692..109Z, he2011propagation, Dresing2012SoPh, 
wang2012effects, qin2013transport, Droge2014JGRA..119.6074D, 
Gomez-Herrero2015ApJ...799...55G, WangAQin15}.
For the second one, the SEP fluxes observed by widely separated  multi-spacecraft  
usually show similar intensities within a small $ \sim 2-3$ factor in different 
positions \citep{Reames1997ApJ...491..414R,McKibben2001ICRC....8.3281M, 
Lario2003AdSpR..32..579L,Tan2009ApJ...701.1753T, Zhang2009ApJ...692..109Z, 
qin2013transport, WangAQin15}.
\cite{McKibben1972JGR....77.3957M} discovered this phenomenon, and 
\cite{Roelof1992GeoRL..19.1243R} named it as ``reservoir''.
Recently, \citet{wang2012effects} use a numerical code (denoted as Shock Particle
Transport Code, SPTC) considering shock as a moving SEP source by 
adopting the model of \citet{Kallenrode1997JGR...10222347K}.
\citet{qin2013transport} reproduced the reservoir phenomenon with 
different shock acceleration efficiency and perpendicular diffusion. 
In addition, \cite{WangAQin15} investigated the spatial and temporal invariance in 
the spectra of gradual SEP events.
In their simulations, SPTC is used with the IMF set as Parker field model, and 
the disturbance of the IMF caused by ICME is ignored. 
They found that shock acceleration efficiency, parallel diffusion, adiabatic cooling, 
and perpendicular diffusion are four important factors in forming the reservoir 
phenomenon, and the first three factors are the main factors with the last factor 
being a secondary one.
The peaks of SEP fluxes are mainly controlled by  shock acceleration efficiency and  
parallel diffusion.
And the fluxes decay in the similar ratio due to the effect of adiabatic cooling.
Furthermore,  because of the effect of perpendicular diffusion,  the longitudinal 
gradient in the SEP fluxes, which is observed by spacecraft located at different
locations, is further reduced.
Observationally, the four factors change significantly in different SEP events 
\citep{Kallenrode1996JGR...10124393K,Kallenrode1997JGR...10222347K}, so that 
only in the gradual SEP events when the values of the controlling effect parameters
are appropriate can the reservoir phenomenon be formed.

Generally, SEP acceleration by shocks are calculated in two major approaches:  in
the first approach, SEPs are  injected at the shock with an assumed injection 
strength \citep{Heras1992ApJ...391..359H,Heras1995ApJ...445..497H, 
Kallenrode1997JGR...10222311K,Lario1998ApJ...509..415L, 
Kallenrode2001JGR...10624989K,wang2012effects,qin2013transport}, while in the
second approach the acceleration of SEPs by shocks are included
\citep{Lee1983JGR....88.6109L,Gordon1999JGR...10428263G,zuo2011,
zuo2013acceleration}.   
Each approach has its own advantages, the first one can provide a reasonable 
description of the SEP fluxes in gradual events by focusing on the transport of 
energetic particles without the thorough knowledge of the details of shock 
acceleration, while the second one can help us to better understand diffusive shock 
acceleration by including more physics details.
{\bf{There are also some studies attempting to combine the advantages of the 
two approaches. 
The SEPs acceleration and transport processes are included in these studies 
\citep{ Ng1999GeoRL..26.2145N,
Zank2000JGR...10525079Z,Li2003JGRA..108.1082L, Rice2003JGRA..108.1369R}.}}

Because the calculation of SEP flux needs a precise mechanism for particle's 
injection into the diffusive shock acceleration, which is currently not available, 
we adopt the first approach to inject SEPs at the shock with an assumed injection 
strength, so that we could focus on interplanetary shock accelerated particles' transport \citep{wang2012effects,qin2013transport,WangAQin15}. 
{\bf{In this paper, as a continuation of our previous research, we further study the 
the shock acceleration efficiency and SEPs transport in a gradual SEP event by comparing our simulations with multi-spacecraft observations.}}
We describe the SEP transport model and the shock model in section 2. 
We show the observation results in section 3. 
We show the shock geometry and the effect of perpendicular diffusion in Section 4. 
We show the simulation results and their comparison with multi-spacecraft 
observations in Section 5. We summary our results in Section 6.


\section{MODEL}
In this work we follow previous research \citep[e.g.,][]{Qin2006JGRA..11108101Q,Qin2011ApJ73828Q,qin2013transport, Zhang2009ApJ...692..109Z,droge2010ApJ, zuo2011,zuo2013acceleration,wang2012effects,Wang2014ApJ789157W}, to model the transport of SEPs.
One can write a three-dimensional focused transport equation as \citep{Skilling1971ApJ...170..265S,schlickeiser2002cosmic, Qin2006JGRA..11108101Q,Zhang2009ApJ...692..109Z}
\begin{eqnarray}
  \frac{{\partial f}}{{\partial t}} = \nabla\cdot\left( \bm
  {\kappa_\bot}
\cdot\nabla f\right)- \left(v\mu \bm{\mathop b\limits^ \wedge}
+ \bm{V}^{sw}\right)
\cdot \nabla f + \frac{\partial }{{\partial \mu }}\left(D_
{\mu \mu }
\frac{{\partial f}}{{\partial \mu }}\right) \nonumber \\
  + p\left[ {\frac{{1 - \mu ^2 }}{2}\left( {\nabla  \cdot \bm{V}^
  {sw}  -
\bm{\mathop b\limits^ \wedge  \mathop b\limits^ \wedge } :\nabla
\bm{V}^{sw} } \right) +
\mu ^2 \bm{\mathop b\limits^ \wedge  \mathop b\limits^ \wedge}  :
\nabla \bm{V}^{sw} }
\right]\frac{{\partial f}}{{\partial p}} \nonumber \\
  - \frac{{1 - \mu ^2 }}{2}\left[ { - \frac{v}{L} + \mu \left
  ( {\nabla  \cdot
\bm{V}^{sw}  - 3\bm{\mathop b\limits^ \wedge  \mathop b\limits^
\wedge}  :\nabla \bm{V}^{sw} }
\right)} \right]\frac{{\partial f}}{{\partial \mu }},\label{dfdt}
\end{eqnarray}
where $f(\bm{x},\mu,p,t)$ is the distribution function averaged over gyrophase, 
$t$ is the time,
$\bm{x}$ is the position in a non-rotating heliographic coordinate system, $p$, $v$, and $\mu$ are the particle momentum, speed, and pitch-angle cosine, respectively, in the solar wind frame, $\bm{\mathop b\limits^ \wedge}$ is a unit vector along the local magnetic field,
$\bm{V}^{sw}=V^{sw}\bm{\mathop r\limits^ \wedge}$ is the solar wind velocity, and the magnetic focusing length $L$ is obtained from the 
divergence of the IMF background strength $B_0$, i.e., 
$L=\left(\bm{\mathop b\limits^ \wedge}\cdot\nabla\\{ln} B_0\right)^{-1}$. 
Here, we use the Parker field model for the IMF.
The equation (\ref{dfdt}) includes almost all important particle transport effects,
such as particle streaming along the field line, adiabatic cooling in the expanding 
solar wind, magnetic focusing in the diverging IMF, and the parallel and 
perpendicular diffusion coefficients.

{\bf{We define a pitch angle diffusion coefficient following}}
\citep{Beeck1986ApJ...311..437B, qin2005model, Qin2006JGRA..11108101Q}
\begin{equation}
{D_{\mu \mu }}(\mu ) = {D_0}v{(R_L k_{min})^{s - 2}}\left({\mu ^{s - 1}} + 
h\right)\left(1 - {\mu ^2}\right),
\end{equation}
where the constant ${D_0}$ is adopted from \cite{Teufel2003A&A397}
\begin{equation}
{D_0} = {\left( {\frac{{\delta {B_{slab}}}}{{{B_0}}}} \right)^2}\frac{{\pi (s - 
1)}}{{4s}}{k_{\min }},
\end{equation}
here  $ \delta {B_{slab}}/B_0 $ is  the magnetic turbulence level of slab component,
$ R_L = pc/\left( {\left| q \right|{B_0}} \right) $ is the 
maximum particle Larmor radius, $q$ is the  particle charge,
$l_{slab}$ is the slab turbulence correlation length,
$ {k_{\min }}=1/l_{slab} $ is the lower limit of wave number of the inertial range in the slab turbulence power spectrum, 
and $s = 5/3$ is the Kolmogorov spectral index of the magnetic field turbulence in the 
inertial range. 
The constant $h$ arises from the non-linear effect of magnetic turbulence on the 
pitch angle diffusion at $\mu=0$ \citep{qin2009pitch,qin2014detailed}.
In the following simulations, we set  $ h=0.01$,  and ${k_{\min }}= 32$ AU$^{-1}$.

Following \citet{Jokipii1966ApJ...146..480J}, \citet{hasselmann1968scattering}, and 
\citet{earl1974diffusive}, the relationship between $D_{\mu \mu }$ and parallel 
mean free path (MFP) $\lambda _\parallel$ is written as
\begin{equation}
\lambda _\parallel   = \frac{{3\upsilon}}{8}\int_{ - 1}^{ + 1}
{\frac{{(1 - \mu ^2 )^2 }}{{D_{\mu \mu } }}d\mu
}.\label{lambda_parallel_1}
\end{equation}
In addition, the parallel diffusion coefficient $\kappa_\parallel$ is related to 
$\lambda_\parallel$ by $\kappa_\parallel=v\lambda_\parallel/3$.

The perpendicular diffusion coefficient is set by using the nonlinear guiding center 
theory \citep{Matthaeus2003ApJ...590L..53M} with the following analytical 
approximation \citep{Shalchi2004ApJ...616..617S,Shalchi2010Ap&SS.325...99S}
\begin{equation}
{\bm{\kappa} _ \bot } = \frac{1}{3}v{\left[ {{{\left( {\frac{{\delta {B_{2D}}}}{{{B_0}}}} \right)}^2}\sqrt {3\pi } \frac{{s - 1}}{{2s}}\frac{{\Gamma \left( {\frac{s}{2} + 1} \right)}}{{\Gamma \left( {\frac{s}{2} + \frac{1}{2}} \right)}}{l_{2D}}} \right]^{2/3}}{\lambda _\parallel }^{1/3}\left( {{\bf{I}} - \bm{ \mathop b\limits^ \wedge} \bm{ \mathop b\limits^ \wedge}  } \right)  
\end{equation}
where $B_{2D}/B_0$  are the turbulence level of 2D component,
and $ l_{2D} $ is the correlation length.
$ \Gamma $ is the gamma function.
$\bm{{\rm I}}$ is a unit tensor. In our simulations,
${l_{2D }}$ is set to $3.1\times 10^{-3}$ AU, ${\left( {\delta {B_{2D}}} 
\right)^2}/{\left( { \delta {B_{slab}}} \right)^2}= 4$, and $s = 5/3$. As a result,
  the values of parallel and perpendicular diffusion coefficients can be altered by 
changing the magnetic turbulence level $\delta B/B_0\equiv
\sqrt{\delta B_{slab}^2+\delta B_{2D}^2}/B_0$.

{\bf{The particle injection into the shock at position ($r,\theta,\varphi$) and time $t$ with momentum $p$ is specified by boundary values from}} 
\citet{Kallenrode1997JGR...10222311K,Kallenrode2001JGR...10624989K,wang2012effects},
and \citet{qin2013transport}
\begin{eqnarray}
 f_b (r,\theta ,\varphi ,p,t) &=& a \cdot \delta (r - \upsilon_s t) \cdot S(r,\theta ,\varphi )
 \cdot p^{  \gamma }  \cdot \xi (\theta ,\varphi ) \nonumber\\
S(r,\theta ,\varphi ){\rm{ }} &=& {\left( {\frac{r}{{{r_c}}}} \right)^{ - \alpha }}\exp \left[ { - \frac{{\left| {\phi (\theta ,\varphi )} \right|}}{{{\phi _c}}}} \right] \nonumber\\
 \xi (\theta ,\varphi ) &=& \left\{ \begin{array}{l}
 1{\kern 21pt}  {\rm{if}}\ \left| {\phi (\theta ,\varphi )}
 \right| \le \phi _s  \\
 0{\kern 21pt}  {\rm{otherwise,}} \\
 \end{array} \right.\label{f_b}
\end{eqnarray}
where $r$ is the solar radial distance, $\upsilon_s$ is the shock speed, 
$\upsilon_s t=r_{b0} + n \cdot \Delta r$ with $n =0,1,2, \cdot \cdot \cdot, n_0$ and
 $\Delta r$ being space interval between two `fresh' injections, $r_{b0}$ is the
inner boundary, $S$ is the shock acceleration efficiency which specifies the particles
 ejection and changes with a power law in radial distance and is exponential towards
 the flank of shock, $r_c$ is the radial normalization parameter, $\alpha$ and 
$\phi_c$ are the shock acceleration efficiency parameters, $\xi$ determines the 
spatial scale of shock front, ${\phi}$ is the angle between the center of shock and 
the point at  the shock front where the particles injected, and ${\phi_s}$ is the 
half width of the shock. $\gamma$ is the energy spectral index of SEP source.

In order to numerically solve the transport equation (\ref{dfdt}), a time-backward 
Markov stochastic process method is used by following 
\citet{Zhang1999ApJ...513..409Z}, see also \citet{Qin2006JGRA..11108101Q} for 
details of the application of the methods to study SEPs. 
Here, we denote the numerical code to calculate the propagation of energetic 
particles, which treats the CME driven shock as a moving particle source, as Shock 
Particle Transport Code, i.e., SPTC.

\section{OBSERVATIONS}

\begin{figure}
\epsscale{0.8} \plotone{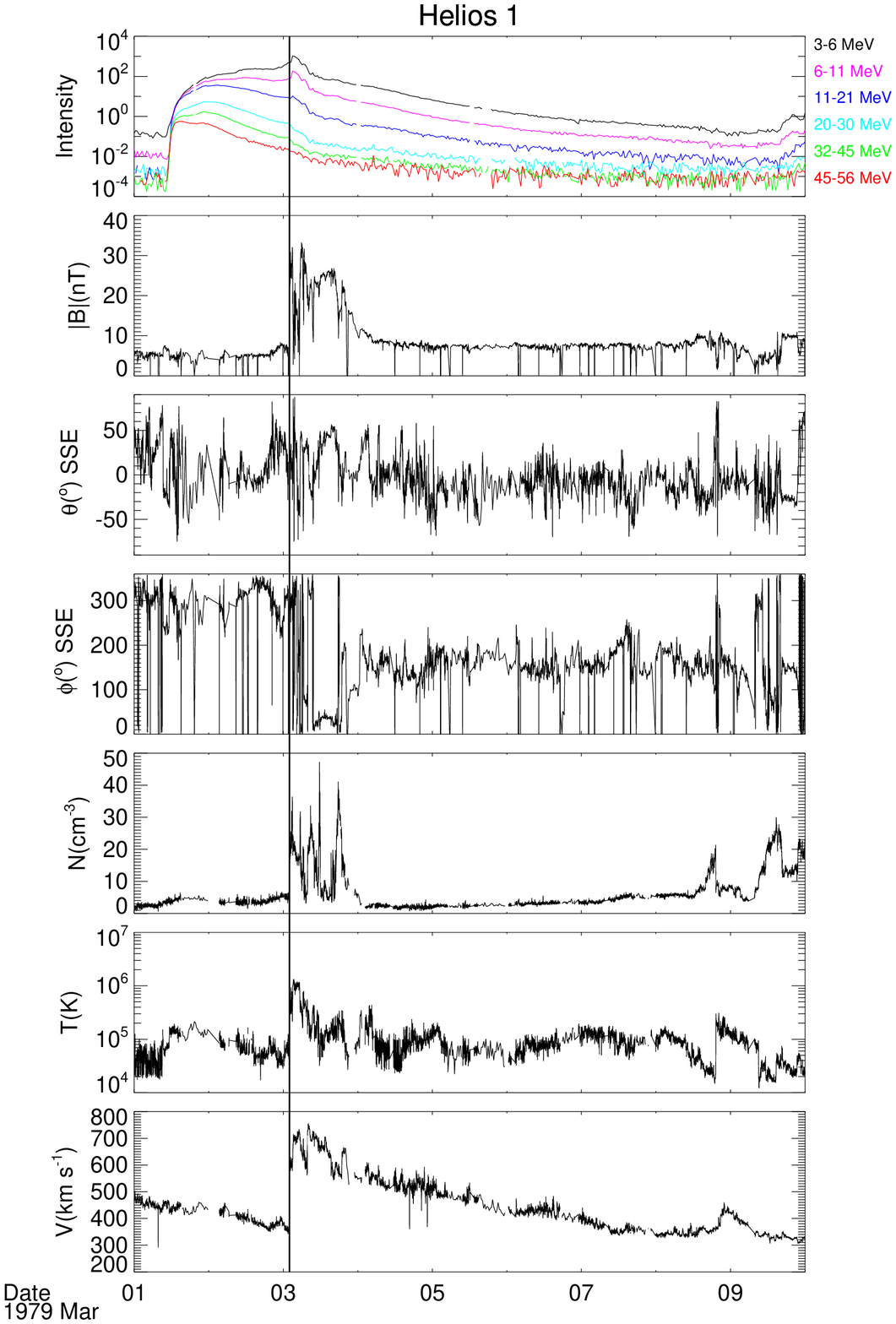} 
\caption{Solar wind plasma and magnetic field parameters observed at the Helios 1 spacecraft. 
From top to bottom, magnetic field strength $\left| {\bf{B}} \right|$, polar and azimuthal field angles $\theta$ and $\phi$  (in SSE coordinate), plasma density $N$, plasma temperature $T$, and 
bulk solar wind speed $V$.
The vertical line indicates a interplanetary shock passage.
\label{Helios1_Plasma}}
\end{figure}

\begin{figure}
\epsscale{0.8} \plotone{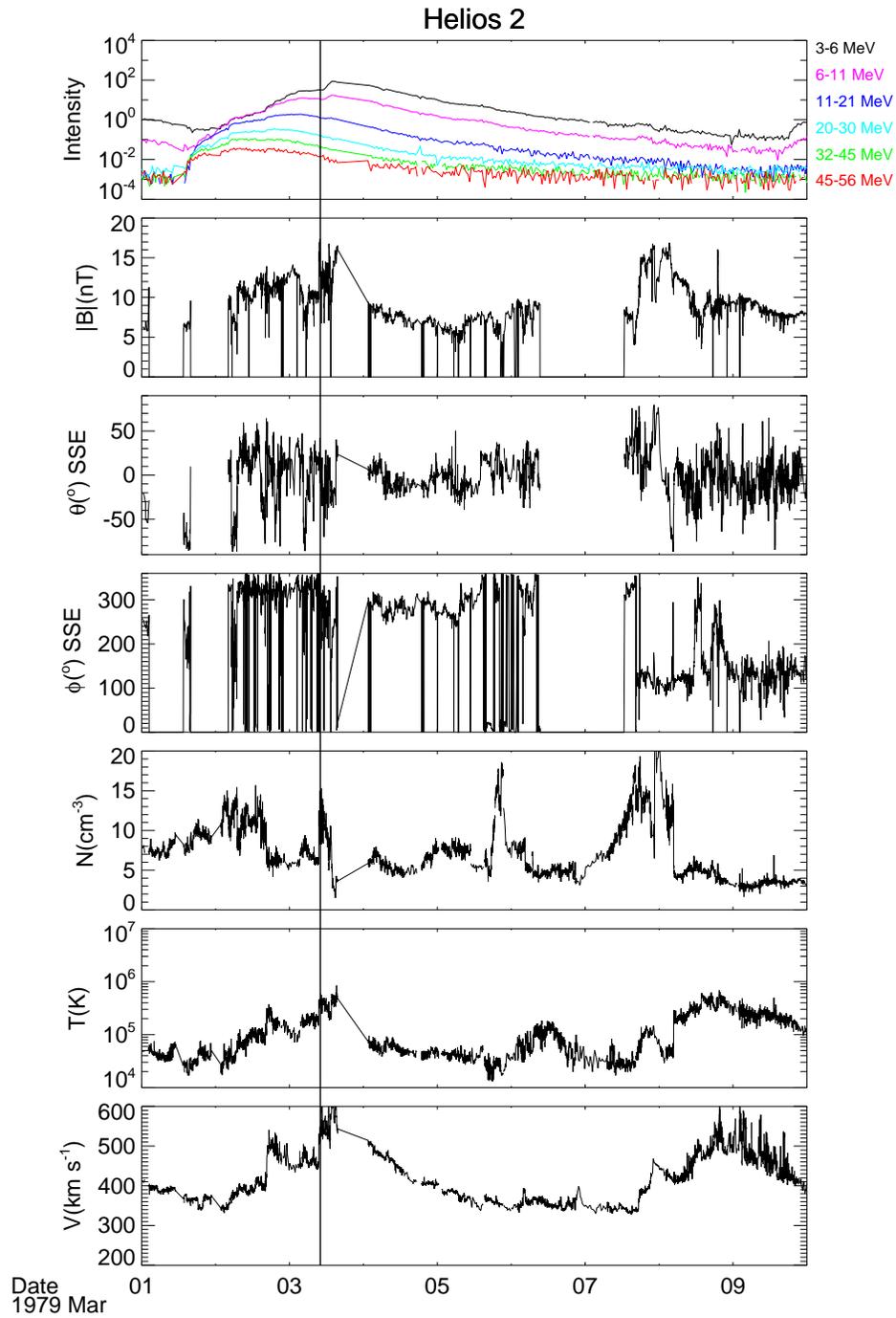} 
\caption{Same as Figure \ref{Helios1_Plasma} except that the parameters are observed at the Helios 2 spacecraft.
\label{Helios2_Plasma}}
\end{figure}

\begin{figure}
\epsscale{0.8} \plotone{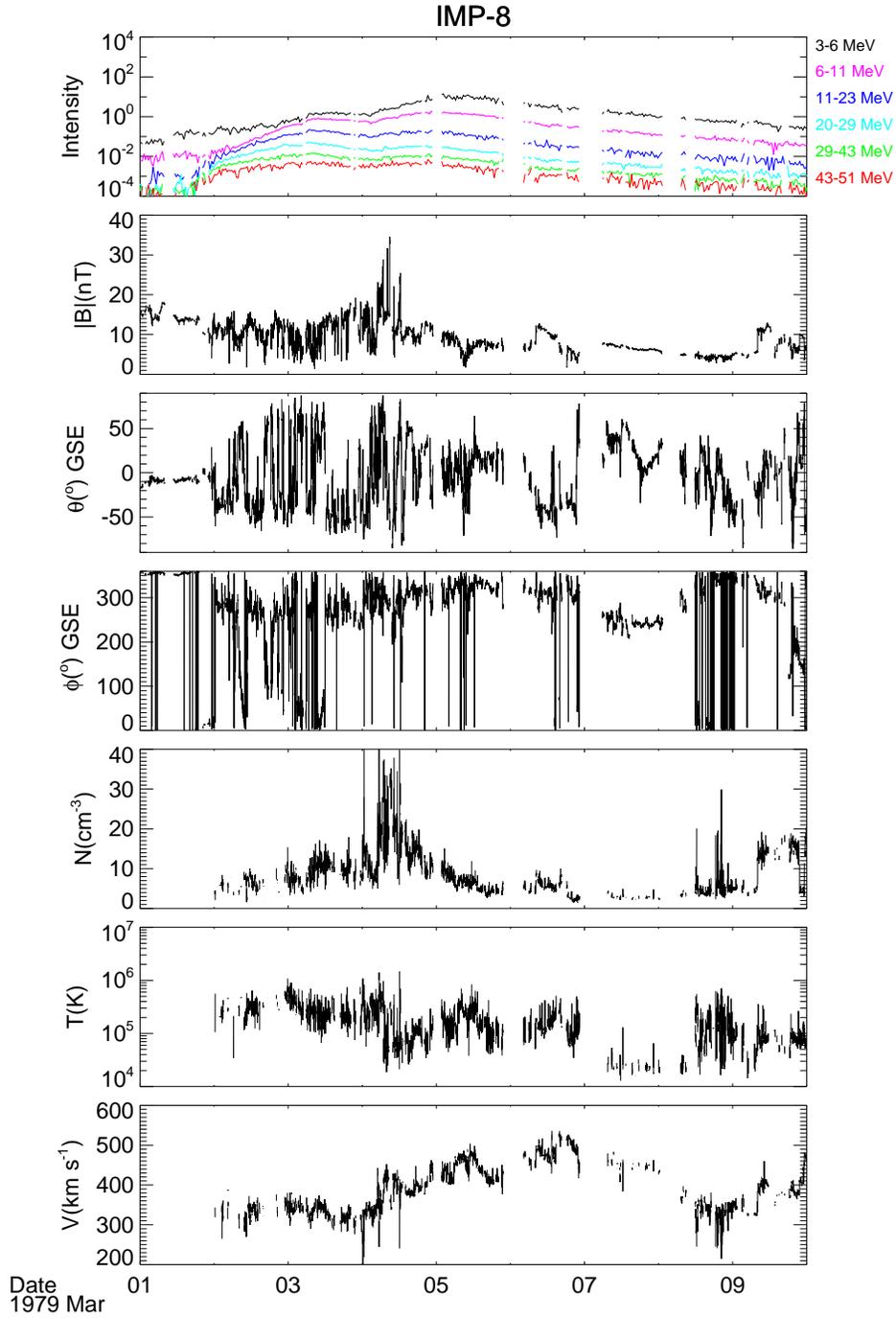} 
\caption{Same as Figure \ref{Helios1_Plasma} except that the parameters are observed at the $IMP$ 8 spacecraft and olar and azimuthal field angles are in GSE coordinate.
\label{Imp8_Plasma}}
\end{figure}

{\bf{The SEP data from the $Helios$  University of Kiel particle
instrument  provides proton and heavier nuclei with energy range between 
$1.3$ and several hundred MeV/nuc, and electrons of  $0.3-0.8$ MeV.   
The SEP data from the $IMP$ $8$ Goddard Medium Energy (GME) Experiment
provides an energy range of $0.5-450$ MeV proton and 
$2-450$ MeV/nuc heavier nuclei,  and relativistic electrons.}}
During the time period March 1-11, 1979, a gradual SEP event was 
observed by $Helios$ 1 and 2, and $IMP$ 8, which were located near $1$ AU ecliptic, 
but at different  longitudes. 
{\bf{On March 1, 1979, the radial distances of $Helios$ 1, $Helios$ 2, and $IMP$ 8 on 
March 1, 1979 are $0.95$ AU, $0.93$ AU, and $0.99$ AU, respectively.
And the $Helios$ 1, $Helios$ 2, and $IMP$ 8
were located at $17^\circ$, $57^\circ$, and $84^\circ$ in the heliographic inertial 
coordinate, respectively.}}
Figures \ref{Helios1_Plasma}, \ref{Helios2_Plasma}, and \ref{Imp8_Plasma} show the 
time series of SEP fluxes, interplanetary magnetic field, and solar wind 
measurements from the $Helios$ 1, $Helios$ 2, and $IMP$ 8, respectively. 
In each of the three figures, from top to bottom, the observation data are for 
SEP fluxes, magnetic field strength $\left| {\bf{B}} \right|$, 
polar and azimuthal field angles 
$\theta$ and $\phi$  in the Selenocentric Solar Ecliptic (SSE) coordinate for $Helios$
1 and 2 but in the Geocentric Solar Ecliptic (GSE) coordinate for $IMP$ 8, plasma
density $N$, plasma temperature $T$, and bulk solar wind speed $V$.
The vertical line indicates a interplanetary shock passage.
From Figure \ref{Helios1_Plasma} we find that during the period March 3-5 $Helios$ 1
observed both a shock and an ICME event characterized by increases in the solar wind
speed, density, temperature, and the rotation of the azimuthal field angle $\phi$. 
From Figure \ref{Helios2_Plasma} we find that $Helios$ $2$ only observed a shock 
without ICME. 
{\bf{The ICME driven shock was detected by the $Helios$ 1 and $Helios$ 2 during
March 3. If the same shock was detected by $IMP$ 8, it should be detected nearly 
in the same time interval. However, during March 3, we find that $IMP$ 8  
observed neither a shock nor an ICME in Figure \ref{Imp8_Plasma}.
This SEP event is also investigated by \cite{Lario2006ApJ...653.1531L} and 
\cite{Reames2010SoPh..265..187R}. In these two studies, they also found
that $IMP$ 8 did not observe a shock.}}

Just before the onset of the March 1-11, 1979 gradual SEP event at $1$ AU, a solar 
flare is observed \citep{Kallenrode1992ApJ...391..370K}. 
The flare is located at S$23$E$58$, and the time of the maximum of soft X-ray is 
at 10:19 on March 1.
If we assume the time of the maximum of soft X-ray to be the moment of
formation of the CME driven shock, the average speed of the shock can be obtained, 
i.e., $0.57$ AU/day and $0.47$ AU/day for the average speed of shocks detected by 
$Helios$ 1 and 2, respectively. 
In this work, we further take an average of $0.57$ AU/day and $0.47$ AU/day, and set
the shock speed to be a constant $0.52$ AU/day.

\section{SHOCK'S GEOMETRY AND PERPENDICULAR DIFFUSION}

{\bf{In fact, the geometry of an interplanetary shock would be very complicate. 
For simplicity, we do not include shock angle (obliquity) in our model, but only
use a cone to model the shock's geometry in our simulations. 
The shock nose (the center of the cone) is located in ecliptic, and the longitude 
of the shock nose 
is set as the same as that of solar flare. In this event, the shock nose points 
to E$58$. Since the locations of spacecraft are known, then we can determine the
relative positions between the shock nose and the spacecraft.
In this case, the $Helios$ 1, $Helios$ 2, and $IMP$ 8 are $9^\circ$ east, 
$31^\circ$ west, and $58^\circ$ west of the shock nose, respectively.
Although the direction of the shock nose is a hypothesis, 
we can use the spacecraft's  in-situ observation to identify 
if the hypothesis is reasonable.
In this SEP event, $Helios$ 1 detected the shock and ICME, 
but $IMP$ 8 didn't. These observations implied that the $Helios$ 1 is located 
the nearest to the shock nose, and $IMP$ 8  is the farthest.
Therefore, the locations of spacecraft in our model are consistent 
with the in-situ observations of spacecraft.  
Furthermore, the solid angle of shock front is set as $35^\circ$, which agrees with 
the fact that both $Helios$ 1 and $Helios$ 2 detected shock, but $IMP$ 8 did not.
According to the observations of $Helios$ 1, there is no significant difference 
in the SEP fluxes when the spacecraft is inside and outside the ICME. 
As a result, the disturbances of IMF caused by ICME could be ignored, and the IMF 
is set as Parker spiral in our simulations.}}

\begin{figure}
\epsscale{0.8} \plotone{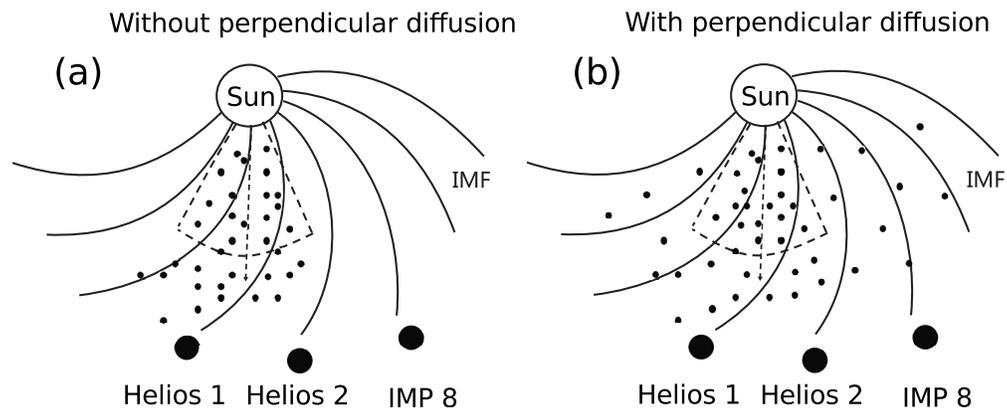} 
\caption{Geometry of a shock cross-section with three spacecraft at different locations. The dashed-arrow radial line indicates the center of the shock. 
The big solid circles represent the three spacecraft, and the small ones represent protons of SEPs.
\label{heliosAndIMP8WithParticles}}
\end{figure}

Based on our shock model and the observations of 
$Helios$ 1, $Helios$ 2 , and $IMP$ 8, we plot a cartoon
for illustrating the cross-section of the shock and the locations of three 
spacecraft in  Figure \ref{heliosAndIMP8WithParticles}. 
In this figure, the shock front is indicated by the dashed arc line, and the shock 
nose is indicated by the dashed-arrow radial line 
passing through the center of the shock. 
The big solid circles indicate the locations of the three observers, $Helios$ 
1 and 2, and $IMP$ 8, and the small ones indicate the protons of SEPs.
As the shock propagates outward, the shock front is connected with the $Helios$ 1 
by the IMF first, then with the $Helios$ 2, at last with the $IMP$ 8.
Note that the spacecraft's field line is not connected to the shock front all the 
time. For example, the field line of the $IMP$ 8 is not connected to the shock at 
the beginning, then as the shock front moves to a larger radial distance than that 
of the $IMP$ 8, the field line becomes connected to the shock. 
Eventually, the observer will be disconnected from the shock as the shock continues 
to propagate outward.

In Figure \ref{heliosAndIMP8WithParticles} (a), we assume that SEPs propagate in the  interplanetary space without perpendicular diffusion. 
In this case, the SEPs can only propagate along the IMF. 
The particles can be detected at the onset time when the spacecraft's IMF is connected to the shock front. 
In Figure \ref{heliosAndIMP8WithParticles} (b), however, we assume that SEPs propagate in the interplanetary space with perpendicular diffusion. 
In this case, SEPs  can cross magnetic field lines. 
The particles can be detected before the spacecraft is connected to the shock by 
field lines, and they can more easily spread in the interplanetary space.
In March 1, 1979 SEP event, the onset time of SEP fluxes observed by three 
spacecraft was very close. Therefore, in order to reproduce the observations with 
our model, perpendicular diffusion must be included in the following simulations.

\section{THE RESULTS OF SEP SIMULATIONS AND THEIR COMPARISONS WITH OBSERVATIONS}

The parameters used in the simulations are listed in tables \ref{paratable} 
and \ref{paratable2}.
The parameters in table \ref{paratable} are the same in all of following 
simulations, but the ones in table \ref{paratable2} are different in individual 
simulations. 
{\bf{In the observations, the speed of solar wind is always changing with 
time, but the variance in the speed would not change our 
main conclusions significantly. As result, we used a constant speed of
$400$ km/s in our simulations.}}

\subsection{Effect of Perpendicular Diffusion on SEP Flux}
\begin{figure}
\epsscale{0.8} \plotone{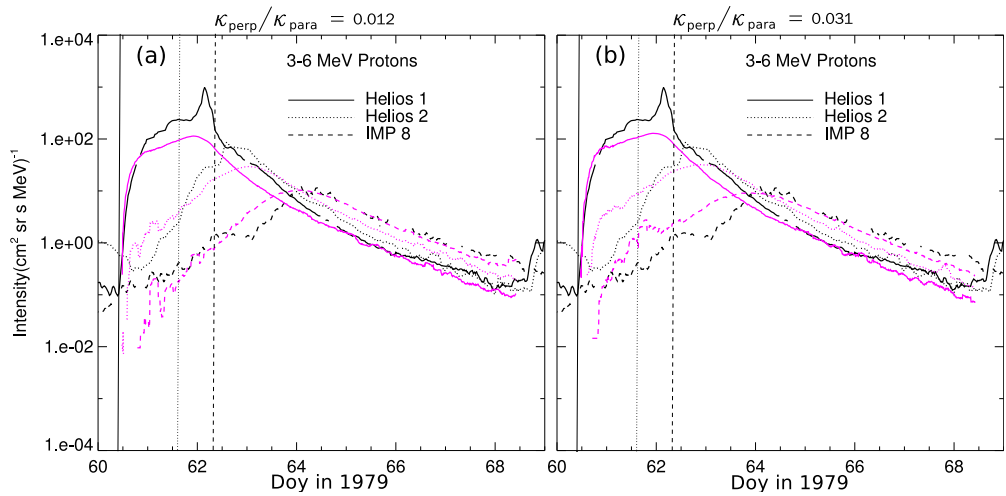} 
\caption{ Comparison of the observations of $3-6$ MeV proton fluxes with black lines and the simulation results of $5$ MeV protons with red lines. 
The observers are located at $1$ AU in the ecliptic, but at different longitudes. 
The black lines show the time profiles of the observation fluxes, and the red lines indicate the simulation fluxes.
The vertical lines indicate the moment of the spacecraft's field line is connected to the shock.
\label{HeliosIMP8ProtonDifferentDiffusion}}
\end{figure}

Figure \ref{HeliosIMP8ProtonDifferentDiffusion} shows the simulation and 
observation results of the time profiles  of SEP fluxes.
The black and red lines indicate the observations of $3-6$ MeV protons and the 
simulations of $5$ MeV protons, respectively.
And the solid, dotted, and dashed lines are corresponding to the observations of 
$Helios$ 1, $Helios$ 2, and $IMP$ 8, respectively. 
The vertical lines indicate the moment when the spacecraft's field lines are 
connected to the shock.
Due to the different locations of the three spacecraft and the width of the shock, 
the spacecraft's field lines are connected to the shock front at different time.
The $Helios$ 1 is connected to the shock by IMF first, then is the $Helios$ 2, at 
last is the $IMP$ 8. The simulations in Figure 
\ref{HeliosIMP8ProtonDifferentDiffusion} (a) are noted as Case 1. 
Note that the parameters for different cases are shown in Table \ref{paratable2}.
The onset time when the shock is connected to $IMP$ 8 by IMF is nearly two days 
later than that to $Helios$ 1. 
With perpendicular diffusion, particles can cross magnetic field lines in the 
interplanetary space, so they can be detected before the spacecraft's magnetic field
line is connected to the shock. 
As a result, the onset time of SEP fluxes in the simulations is very close because 
of the effect of perpendicular diffusion. 
As we can see, the simulations roughly agree with the observations in Figure \ref
{HeliosIMP8ProtonDifferentDiffusion} (a).
The simulations in Figure \ref{HeliosIMP8ProtonDifferentDiffusion} (b) are noted as
Case 2. 
Figure \ref{HeliosIMP8ProtonDifferentDiffusion} (b) is the same as Figure 
\ref{HeliosIMP8ProtonDifferentDiffusion} (a) except that the ratio of perpendicular 
diffusion coefficient to parallel one is larger.  
In Figure \ref{HeliosIMP8ProtonDifferentDiffusion} (b), SEP flux observed by 
$Helios$ 1 is very close to that in  Figures 
\ref{HeliosIMP8ProtonDifferentDiffusion} (a). 
However, during the early stage of flux rising phase, SEP fluxes observed by 
$Helios$ 2 and  $IMP$ 8 in Figure \ref{HeliosIMP8ProtonDifferentDiffusion} (b) are 
slightly larger than that in Figure \ref{HeliosIMP8ProtonDifferentDiffusion} (a).
This is because all particles arrive at $Helios$ 2 and  $IMP$ 8 by crossing field 
lines before the observers field lines are connected to the shock front.
Due to the effect of stronger perpendicular diffusion in Figure 
\ref{HeliosIMP8ProtonDifferentDiffusion} (b) than that in  Figure 
\ref{HeliosIMP8ProtonDifferentDiffusion} (a), during the decay phases, the spatial 
gradients of SEP fluxes in Figure \ref{HeliosIMP8ProtonDifferentDiffusion} (b) are 
slightly smaller than that in Figure \ref{HeliosIMP8ProtonDifferentDiffusion} (a). 
In this SEP event, the effect of perpendicular diffusion is very important when the 
spacecraft's field line is disconnected from the shock front, and it also helps to
reduce the spatial gradient in the decay phase of SEP fluxes.

\subsection{Effect of Source Injection Profile on SEP Flux}
\begin{figure}
\epsscale{0.8} \plotone{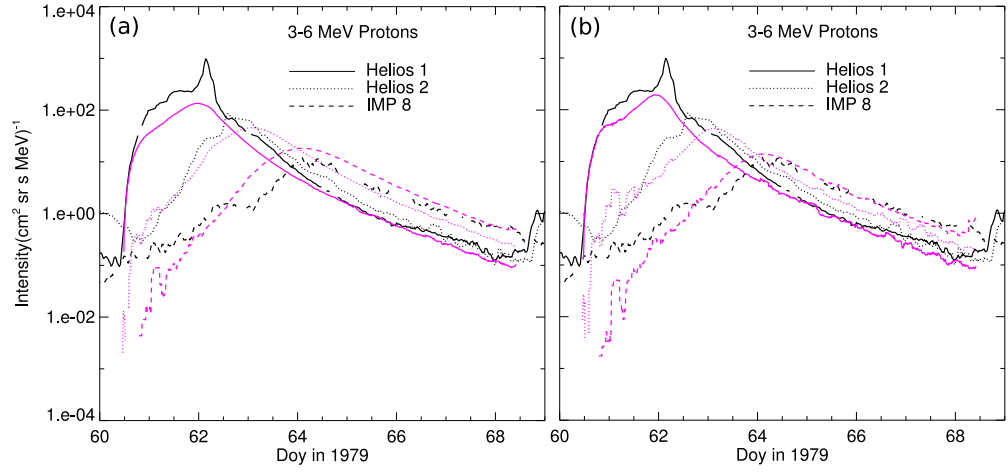} 
\caption{ Same as Figure \ref{HeliosIMP8ProtonDifferentDiffusion} (a) but with  different shock acceleration efficiency parameters $\alpha$ and ${\phi _c}$.
\label{HeliosIMP8Proton_DifferentShockParameters}}
\end{figure}

Figures \ref{HeliosIMP8Proton_DifferentShockParameters}  (a) and (b) are the same as
Figure \ref{HeliosIMP8ProtonDifferentDiffusion} (a) but with different shock 
acceleration efficiency parameters, and the simulations in Figures 
\ref{HeliosIMP8Proton_DifferentShockParameters}  (a) and (b) are noted as Case 3
and Case 4, respectively. 
In simulations of Figure \ref{HeliosIMP8Proton_DifferentShockParameters} (a), the 
shock acceleration parameter $\alpha$ is set to $2.5$.
Comparing with the simulations in  Figure \ref{HeliosIMP8ProtonDifferentDiffusion} 
(a), the shock acceleration efficiency decreases more slowly with radial distance, and
 the longitudinal gradient in the simulation fluxes is larger.
In the simulations of Figure \ref{HeliosIMP8Proton_DifferentShockParameters} (b), 
the shock acceleration parameter ${\phi _c}$ is set to $5^\circ$.
Comparing with the simulations in Figure \ref{HeliosIMP8ProtonDifferentDiffusion} 
(a), the shock acceleration efficiency decreases more quickly toward the shock flank,
 and the simulation fluxes increase more quickly during the rising phase.
The peak intensity of SEP fluxes are mainly determined by shock acceleration 
efficiency, and the gradient of SEP fluxes in the decay phase is more sensitive
 to the shock 
acceleration efficiency parameters than the perpendicular diffusion coefficient.
In this SEP event, at the peak time of flux for $Helios$ $2$ ($IMP$ $8$), the flux 
for $Helios$ $2$ ($IMP$ $8$) was similar to that for $Helios$ $1$ which had been in
decay phase. 
The reservoir phenomenon is formed in the decay phase in the SEP fluxes.

\subsection{Effect of Adiabatic Cooling on SEP Flux}
\begin{figure}
\epsscale{0.5} \plotone{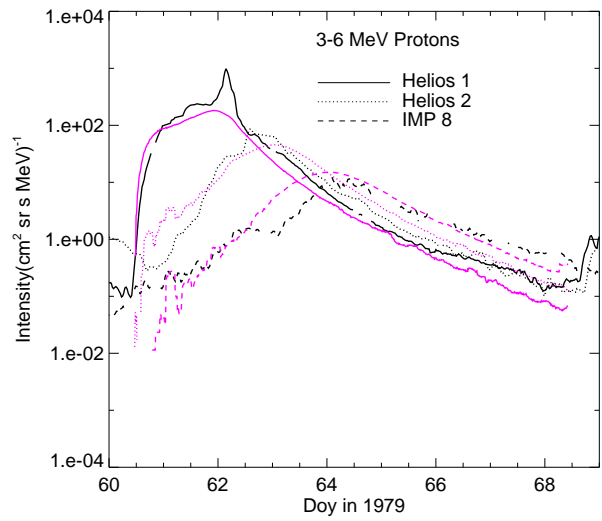} 
\caption{ Same as Figure \ref{HeliosIMP8ProtonDifferentDiffusion} (a) but with  different injection spectrum $\gamma$.
\label{HeliosIMP8Proton_DifferentGamma}}
\end{figure}

Figures \ref{HeliosIMP8Proton_DifferentGamma}  is the same as Figure 
\ref{HeliosIMP8ProtonDifferentDiffusion} (a) but with different injection spectrum. 
The simulations in Figure \ref
{HeliosIMP8Proton_DifferentGamma} is noted as Case 5.
When propagating in the heliosphere, SEPs would lose energy because of the effect of
 adiabatic cooling, so that the SEPs with the same energy observed by spacecraft 
later generally originate with higher energies at the source.
Because the source has a negative energy spectral index $\gamma$, the flux decreases
more quickly with a smaller $\gamma$.
The fluxes observed by $Helios$ 1, $Helios$ 2, and $IMP$ 8  decay as a similar ratio
due to the effect of adiabatic cooling.
In the simulations of Figure \ref{HeliosIMP8Proton_DifferentGamma}, the injection 
spectral index $\gamma$ is set to $-8.5$.
Comparing with the simulations in  Figure \ref{HeliosIMP8ProtonDifferentDiffusion} 
(a), the simulation fluxes decrease more quickly, and the decay ratios of SEP fluxes
are mainly determined by the effect of adiabatic cooling and the energy spectral 
index $\gamma$.

\begin{figure}
\epsscale{0.5} \plotone{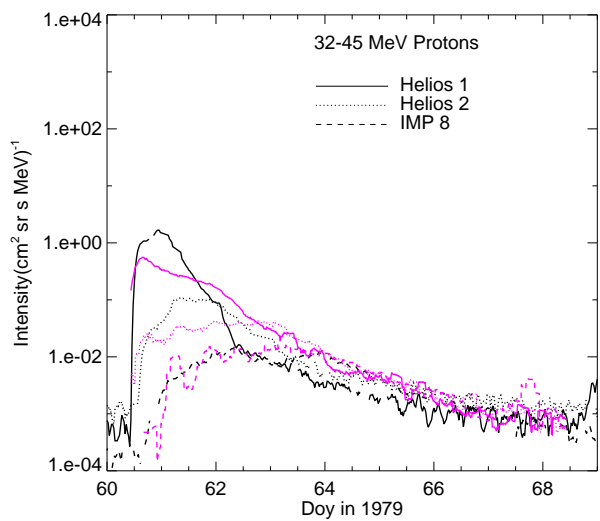} 
\caption{ Comparison of the observations of $32-45$ MeV proton fluxes with black 
lines and the simulation results of $35$ MeV protons with red lines. 
The observers are located at $1$ AU in the ecliptic, but at different longitudes. 
The black lines show the time profiles of the observation fluxes, and the red lines 
indicate the simulation fluxes.
\label{35MeVProtons}}
\end{figure}

{\bf{\subsection{Higher Energetic Particles}
In Figures \ref{35MeVProtons}, the black and red lines indicate the observations of 
$32-45$ MeV protons and the simulations of $35$ MeV protons, respectively.
The simulations in Figure \ref{35MeVProtons} is
noted as Case 6. In simulations of Figure \ref{35MeVProtons}, the shock
acceleration parameter $\alpha$ is set to $4$.
Comparing with the simulations in  Figure \ref{HeliosIMP8ProtonDifferentDiffusion} 
(a), the peak time of simulation fluxes in Figures \ref{35MeVProtons} comes earlier.
This means that source acceleration efficiency decreases with radial distance faster
at higher energy channel.}}

\section{DISCUSSION AND CONCLUSIONS}

In this work, we study a gradual SEP event which was observed by $Helios$ 1 and 2, and $IMP$ 8. 
The event studied in this paper began on March 1, 1979, and lasted nearly eight days. 
By solving a three-dimensional focused transport equation, the fluxes observed by the three spacecraft are calculated.
The transport equation we use in this work includes many important particle transport effects, such as particle streaming along the field line, solar wind convection, adiabatic cooling, magnetic focusing, and the diffusion coefficients parallel and perpendicular to the IMF. 
By comparing the simulations and the observations, we get the shock efficient parameters and diffusion coefficients to get the best fitting. 
The following are our major findings.

The effect of perpendicular diffusion is very important when the spacecraft's field line is disconnected with the shock front.
In 1979/03/01 SEP event, the in-situ observation shows that an ICME is detected by 
$Helios$ 1, but not by $Helios$ 2 or $IMP$ 8 which are in the west of $Helios$ 1.
Therefore, $Helios$ 1 is located near the center of shock front but 
$Helios$ 2 and $IMP$ 8 are located in the west flank of the shock,
if it is assumed that the ICME is located behind the center of the shock.
Furthermore, the interplanetary shock is only observed by $Helios$ 1 and 2, 
but not by $IMP$ 8. 
In the case without perpendicular diffusion, the $IMP$ 8 can not detect any SEP 
when  the radial distance of the shock is smaller than $1$ AU. 
However, according to the observations of $IMP$ 8, the particles can be detected by 
$IMP$ 8 before $IMP$ 8's field line is connected to the shock front. 
In order to reproduce the observations of $3-6$ MeV and $32-45$ MeV protons in the 
simulations, the 
perpendicular diffusion coefficient with the level of about $1\% - 3 \%$ of parallel
one at $1$ AU should be included,

The peaks of fluxes are mainly determined by the shock acceleration efficiency, and 
the decay ratios of SEP fluxes are mainly due to the effect of adiabatic cooling and
 the energy spectral index. Due to the effect of adiabatic cooling, the fluxes 
observed by $Helios$ 1, $Helios$ 2, and $IMP$ 8  decay as a similar ratio.
The shock  acceleration efficiency $S$ is assumed as 
${r^{ - \alpha }}\exp \left( { - \left| \phi  \right|/{\phi _c}} \right)$ in the 
simulations. The shock acceleration efficiency is weaker in the flank of the shock 
front than that in the center, and decreases as the solar radial distance $r$ 
increases. 
{\bf{For $3-6$ MeV protons, by comparing the simulations and the 1979/03/01 
SEP event observed by $Helios$ 1, $Helios$ 2, and $IMP$ 8, we find 
that  $\alpha$ is set to $3$,  $\phi_c$ is  set to 
 $10 ^\circ$,  and the energy spectral index $\gamma$ is set to  $-6.5$.
We also calculated the SEP fluxes in the cases of different model parameters 
which are listed in the  Table \ref{paratable2}.}}
{\bf{For $32-45$ MeV protons,  we find that  $\alpha$ is set to about $4$ in the 
Case 6 of  Table \ref{paratable2}. 
Comparing with the simulations in the Case 1 of $3-6$ MeV protons, 
the $\alpha$ is larger in the case of $32-45$ MeV protons, 
which means the source acceleration efficiency decreases
with radial distance faster at higher energy channel.}}

The longitudinal gradient of SEP fluxes in the decay phase 
observed by different spacecraft is more sensitive to the shock 
acceleration efficiency parameters than the perpendicular diffusion coefficient.
In 1979/03/01 event, at the peak time of flux for $Helios$ 2 ($IMP$ 8), the flux
for $Helios$ 2 ($IMP$ 8) was similar to that for $Helios$ 1 which was in 
decay phase, and the reservoir phenomenon was formed in the decay phase of SEP 
fluxes. 
{\bf{In this event, the reservoir phenomenon is also reproduced in the simulations.}}
However, in some other SEP events, at the peak time of flux for $Helios$ 2 
($IMP$ 8), the flux for $Helios$ 2 ($IMP$ 8) was significantly different than that 
for $Helios$ 1 which was in decay phase. The reservoir phenomenon can not be formed
 in the decay phase of SEP fluxes with normal diffusion coefficients 
 {\bf{( ${\kappa _ \bot }/{\kappa _\parallel }$ is a few percent).}}

The IMF is set to the Parker field model in our simulations, and the  disturbance of
 the IMF caused by  ICME is ignored. 
However, according to the observations of $Helios$ 1 in this event, there is no 
obvious difference in the SEP fluxes after the spacecraft enter the ICME comparing 
to the fluxes detected by $Helios$ 1 when it is out of ICME. 
As a result, we assume that the disturbance of the IMF would not change the main 
results of our simulations. 
{\bf{Furthermore, our model can not be used to study the shock-flare-mixing SEP events.
In the future work, we plan to include the particle source combined with shock and 
flare to model the shock-flare-mixed SEP events.}} And we also
intend to include a realistic three-dimensional ICME shock with 
the disturbance of IMF caused by ICME, so that the SEP acceleration and transport in
 the heliosphere can be investigated more precisely.

\acknowledgments
The authors thank the anonymous referee for valuable comments. 
We are partly supported by grants   NNSFC 41374177,  NNSFC 41125016, and NNSFC 41304135, the CMA grant GYHY201106011, and the Specialized Research Fund for State Key Laboratories of China. 
The computations were performed by Numerical Forecast Modeling R\&D and VR System of State Key Laboratory of Space Weather and Special HPC work stand of Chinese Meridian Project.
We are grateful to the plasma and SEP data provided by the $Helios$ and $IMP$ 8 teams.  

%




\clearpage

\clearpage

\clearpage

\clearpage

\clearpage

\clearpage

\begin{table}
\begin{centering}
\caption {Model Parameters Used in the Calculations.\label{paratable}}
\begin{tabular} {|l|l|l|} \tableline
Parameter & Physical meaning & Value \\
\tableline\tableline
 $V^{sw}$ & solar wind speed & $400$ km/s\\
\tableline
$r_c$ & radial normalization parameter & $0.05$ AU\\
\tableline
$\Delta r$ & shock space interval between two ‘fresh’ injections & $0.001$ AU\\
\tableline
 $v_s$ & shock speed & $0.52$ AU/day\\
\tableline
 $\phi_s$ & shock width & $35^\circ$\\
\tableline
$r_{b0}$ & inner boundary & $0.05$ AU\\
\tableline
$r_{b1}$ & outer boundary & $50$ AU\\
\tableline
\end{tabular}
\end{centering}
\end{table}


\begin{table}
\begin{centering}
\caption {Model Parameters Used in the Calculations.\label{paratable2}}
\begin{tabular} {|c|c|c|c|c|c|c|} \tableline
Case & $\alpha$ & $\phi_c$  & ${\lambda _\parallel}^a$   & 
${{\bf{\kappa }}_ \bot}^a $ & $\gamma$  & 
$\left(\delta B/B_0\right)^2$\\
\tableline\tableline
1  & $3.0$ & $10^\circ$     & $0.48$ AU & $1.2\%  \times {{{\kappa }}_\parallel  }$
  & $-6.5$    & $0.4$  \\
\tableline
2  & $3.0$ & $10^\circ$     & $0.24$ AU & $3.1\%  \times {{{\kappa }}_\parallel  }$  & $-6.5$    & $0.8$   \\
\tableline
3  & $2.5$ & $10^\circ$     & $0.48$ AU & $1.2\%  \times {{{\kappa }}_\parallel  }$  & $-6.5$    & $0.4$   \\
\tableline
4  & $3.0$ & $5^\circ$      & $0.48$ AU & $1.2\%  \times {{{\kappa }}_\parallel  }$  & $-6.5$    & $0.4$   \\
\tableline
5  & $3.0$ & $10^\circ$     & $0.48$ AU & $1.2\%  \times {{{\kappa }}_\parallel  }$  & $-8.5$    & $0.4$   \\
\tableline
6  & $4.0$ & $10^\circ$     & $0.66$ AU & $0.97\%  \times {{{\kappa }}_\parallel  }$  & $-6.5$    & $0.4$   \\
\tableline
\end{tabular}
\end{centering}
\tablenotetext{a }{ for protons in the ecliptic at $1$ AU.}
\end{table}
\clearpage

\end{document}